\documentclass[parindent=0pt]{article}
\usepackage{booktabs}
\usepackage{makecell}
\usepackage{subcaption}

\usepackage[utf8]{inputenc}     
\usepackage[english]{babel}     
\usepackage[a4paper, left=1in, right=1in, top=1.25in, bottom=1.25in]{geometry}
\usepackage{graphicx}           

\usepackage{amsmath,amssymb}    
\usepackage{amsthm}             
\usepackage{mathtools}          
\usepackage{enumitem}           
\usepackage{listings}           
\usepackage{todonotes}          
\usepackage{hyperref}           
\usepackage{float}
\usepackage{amsthm}
\usepackage{amsmath}
\usepackage{amscd}
\usepackage{amssymb}
\usepackage{mathrsfs}
\usepackage{pdfpages}
\usepackage{setspace}
\usepackage{physics}
\usepackage{tcolorbox}

\usepackage{xcolor}
\usepackage{tikz}
\usepackage[toc,page,title,titletoc,header]{appendix}
 \usepackage{url}
 \usepackage{babel} 
\usepackage{csquotes}
\usepackage[f]{esvect}
\usepackage{blindtext}

\newcommand{\R}{\mathbb{R}}

\newcommand{\E}{\mathbb{E}}
\newcommand{\C}{\mathbb{C}}

\newcommand{\g}{\mathfrak{g}}

\newcommand{\oo}{\mathfrak{o}}

\let\phi=\varphi
\let\epsilon=\varepsilon

\theoremstyle{definition}

\newtheorem{result}{Result}

\newtheorem*{definition*}{\textit{Definition}}
\newtheorem*{proposition*}{\textit{Proposition}}

\newcommand{\adots}{\mathinner{\mkern1mu \raise1pt
\hbox{.} \mkern2mu \raise4pt \hbox{.} \mkern2mu \raise7pt
\vbox{\kern7pt \hbox{.}} \mkern1mu}}
\def\[{[\![}
\def\]{]\!]}

\newlist{regimes}{enumerate}{1}
\setlist[regimes]{label={\textbullet\ Regime (\arabic*):}, leftmargin=*}
\usepackage{authblk} 

\title{%
   $\mathcal{R}$-transforms for non-Hermitian matrices: \\
   a spherical integral approach
}

  \author[1,2]{Pierre Bousseyroux\thanks{Email: pierre.bousseyroux@polytechnique.edu}}
\author[3]{Marc Potters}

\affil[1]{Econophysics Lab, Institut Louis Bachelier, 28 Pl. de la Bourse, Palais Brongniart, 75002 Paris, France}
\affil[2]{LadHyX, UMR CNRS 7646, Ecole Polytechnique, Institut Polytechnique de Paris, 91128 Palaiseau, France}
\affil[3]{Capital Fund Management, Paris, France}

\newenvironment{remarks}{
  \par\vspace{1ex}
  \noindent\textbf{Remarks.}\begin{itemize}\setlength\itemsep{0pt}}
  {\end{itemize}\par\vspace{1ex}}

\begin{document}

\maketitle

\begin{abstract}
    In this paper, we establish a connection between the formalism of $\mathcal{R}$-transforms for non-Hermitian random matrices and the framework of spherical integrals, using the replica method. This connection was previously proved in the Hermitian setting \cite{guionnet2005fourier} and in the case of bi-invariant random matrices \cite{benaych2011rectangular}. We show that the $\mathcal{R}$-transforms used in the non-Hermitian context in fact originate from a single scalar function of two variables. This provides a new and transparent way to compute $\mathcal{R}$-transforms, which until now had been known only in restricted cases such as bi-invariant, Hermitian, or elliptic ensembles.
\end{abstract}

\bigskip
\setcounter{footnote}{3} 

Non-Hermitian random matrices arise broadly in physics, information theory, and complex systems, from open quantum systems to neural dynamics \cite{HaakeEtAl1992,FyodorovSommersJETP1996,FyodorovSommersJMP1997,sommers1988spectrum,RajanAbbott2006}. Historically, a key role was played by the \emph{log–determinant potential}
\begin{equation}
\Delta_{\vb{M}}(\omega,z) \;=\; \lim_{N\to +\infty}\frac{1}{N}\,\log\;\!\det\!\bigl(\omega^2 \vb{1} - |\,z \vb{1} - \vb{M}\,|^2\bigr),
\end{equation}where $z\in\mathbb{C}$, $\omega>0$ is a spectral regulator, and $|\,z \vb{1}-\vb{M}\,|^2 := (z \vb{1}-\vb{M})^*(z \vb{1}-\vb{M})$ \footnote{For a matrix $\vb{A}$, we write $\vb{A}^* := \overline{\vb{A}}^{\,T}$ for the conjugate transpose (adjoint).} where $\vb{M}$ is a large non-Hermitian matrix. Introduced by Girko via hermitization in his circular/elliptic laws~\cite{Girko1985_CircularLaw_Eng,girko1986elliptic,sommers1988spectrum}, $\Delta_{\vb{M}}$ became standard in the 1990s physics literature—through replica and supersymmetry techniques—as a generating function for spectral densities and Dyson–Schwinger equations~\cite{feinberg1997non,FeinbergZeeNPB504,janik1997non,JanikEtAlPRE1997,ChalkerWangPRL1997}. In the large-$N$ regime, derivatives of $\Delta_{\vb{M}}$ yield both the eigenvalue density (Brown measure, via $\partial_{z,\bar z}$) and the singular-value resolvent (via $\partial_{\omega}$), thereby unifying eigenvalue and singular-value information in a single potential~\cite{sommers1988spectrum,janik1997non}.

From a mathematical perspective, the same logarithmic potential underlies the \emph{Brown measure}, a rigorous substitute for a spectral distribution of non-normal operators: one considers the singular values of $z \vb{1}-\vb{M}$, forms the associated logarithmic potential, and then applies the two-dimensional Laplacian in $z$ to obtain a probability measure on $\mathbb{C}$. This construction goes back to Brown and was extended by Haagerup and Schultz~\cite{brown1986lidskii,haagerup2007brown}. A useful principle is that the family of singular-value laws of $z \vb{1}-\vb{M}$ (as $z$ varies) encodes exactly the Brown measure~\cite{bordenave2012around}. We also note that the Brown measure can behave subtly—for instance, it may fail to depend continuously on $*$-moments~\cite{Tao2012topics,Anderson2010introduction,Biane1999computation,Sniady2002random}.

A unifying analytical viewpoint is the \emph{electrostatic analogy} on the complex plane, wherein eigenvalues act as charges and a logarithmic potential is governed by a Green’s function whose Laplacian yields the spectral density via Gauss’s law~\cite{sommers1988spectrum,FyodorovSommersJMP1997}.  To control non-Hermitian spectra we adopt the \emph{Hermitian reduction} (a.k.a.\ hermitization) formalism~\cite{janik1997non,feinberg1997non,FeinbergZeeNPB504}. Given an $N\times N$ matrix $\vb{M}$ and spectral parameter $z\in\C$, embed $\vb{M}$ into the $2N\times2N$ block-Hermitian matrix
\begin{equation}
\vb{H}(z)\;:=\;
\begin{pmatrix}
0 & z\vb{1}-\vb{M}\\
\overline{z}\vb{1}-\vb{M}^* & 0
\end{pmatrix}.
\end{equation}
For  $\omega\in\C$, define the block resolvent
\begin{equation}
\vb{G}_{\vb{M}}(\omega,z)\;:=\;\bigl[\omega \vb{1}-\vb{H}(z)\bigr]^{-1}
\;=\;
\begin{pmatrix}
\vb{G}_{11} & \vb{G}_{12}\\
\vb{G}_{21} & \vb{G}_{22}
\end{pmatrix},
\end{equation}
and the block-trace scalars
\begin{equation}
\g_{ij}^{N}(\omega,z)\;:=\;\frac{1}{N}\Tr\,\vb{G}_{ij}(\omega,z),\qquad i, j\in\{1,2\}.
\end{equation}
The \emph{quaternionic} (matrix-valued) resolvent is then defined as
\begin{equation}\label{eq:quaternionic-resolvent}
\mathcal{G}_{\vb{M}}^{N}(\omega,z)
\;:=\;
\begin{pmatrix}
\g_{11}^{N}(\omega,z) & \g_{12}^{N}(\omega,z)\\
\g_{21}^{N}(\omega,z) & \g_{22}^{N}(\omega,z)
\end{pmatrix}.
\end{equation}
Its large-$N$ limit is given by
\begin{equation}\label{defG}
    \mathcal{G}_{\vb{M}}(\omega,z)
    \;=\;
    \lim_{N \to \infty} \mathcal{G}_{\vb{M}}^{N}(\omega,z)
    \;=\;
    \begin{pmatrix}
        \g_1(\omega, z) & \g_2(\omega, z) \\
        \overline{\g_2}(\omega, z) & \g_1(\omega, z)
    \end{pmatrix},
\end{equation}
where
\begin{align}
\g_1(\omega, z)
&=\;
\tau\!\left[\omega\bigl(\omega^2 \vb{1}-(z \vb{1}-\vb{M})(z \vb{1}-\vb{M})^*\bigr)^{-1}\right],
\\[0.5em]
\g_2(\omega, z)
&=\;
-\,\tau\!\left[\bigl(\omega^2 \vb{1}-(z \vb{1}-\vb{M})(z \vb{1}-\vb{M})^*\bigr)^{-1}(z \vb{1}-\vb{M})^*\right].
\end{align}
and $\tau := \lim_{N \to \infty} \frac{1}{N}\,\mathrm{Tr}$ denotes the normalized trace in the large-$N$ limit. Note that in the limit $\omega \to 0$, $\g_2(\omega, z)$ converges to $\g(z)$, which coincides with the Stieltjes transform defined as  
\begin{equation}
    \g(z)
    \;=\;
    \tau\!\left([z\vb{1} - \vb{M}]^{-1}\right).
\end{equation}
Equivalent formulations arise either from direct regularization of the electrostatic potential~\cite{JanikThesis1996,janik1997non,JanikEtAlPRE1997}; these are now known to be interrelated and amenable to a quaternionic-algebraic framework~\cite{JaroszNowakNovel2004}. A rigorous operator-valued formulation and subordination scheme is provided by free probability~\cite{belinschiMaiSpeicher2013,belinschiSniadySpeicher1506}.

Diagrammatic techniques for $\mathcal{G}$ were developed in the late 1990s~\cite{JanikEtAlPRE1997,janik1997non,feinberg1997non,FeinbergZeeNPB504,ChalkerWangPRL1997}. They lead to a \emph{matrix-valued Blue’s function}—the functional inverse of $\mathcal{G}$—which satisfies a simple addition rule for freely independent non-Hermitian matrices (heuristically, in the regulator limit $\omega\to 0$):
\begin{equation}\label{eq:matrix-blue-addition}
\mathcal{B}_{X+Y}(\mathcal{G})
\;=\;
\mathcal{B}_{X}(\mathcal{G})
\;+\;
\mathcal{B}_{Y}(\mathcal{G})
\;-\;
\mathcal{G}^{-1}.
\end{equation}
 This is the non-Hermitian analogue of Voiculescu’s linearization by the $R$-transform and motivates matrix-valued $R$-/Blue transforms~\cite{voiculescu1986addition,voiculescu1992free,speicher1994,NicaSpeicher2006}. In parallel, multiplicative analogues culminate in a non-Hermitian $S$-transform for products of independent random matrices~\cite{BurdaJanikNowakMultiplication}. Throughout, we systematically work in the limit $\omega\to 0$.

Starting from the Blue function, one can define a matrix-valued $R$-transform, expressible either in a $2\times2$ formalism~\cite{BurdaJanikNowakMultiplication} or in a quaternionic formulation~\cite{jarosz2006random}, by
\begin{equation}\label{eq}
\mathcal{R}(\mathcal{G}):=\mathcal{B}_{X}(\mathcal{G}) \;-\; \mathcal{G}^{-1}.
\end{equation}
In practice, fully explicit formulas are rare and are known in a few controlled settings. The elliptic case admits a diagrammatic treatment where the matrix $\mathcal{R}$ takes simple form~\cite{BurdaJanikNowakMultiplication}. The bi-invariant case\footnote{A random matrix $\vb{M}$ is said to be bi-invariant if $\vb{M}$ and $\vb{U}\vb{M}\vb{V}$ have the same distribution for all unitary matrices $\vb{U}$ and $\vb{V}$.} constitutes another tractable regime, in which $\mathcal{R}$ reduces to a single-variable function given by
\begin{equation}
\mathcal{R}(\g_1, \g_2) = R(\g_{1})\vb{I}_2,
\end{equation}
corresponding to the so-called “$R$-diagonal’’ situation in the mathematical literature,
where $\vb{I}_2$ denotes the $2\times2$ identity matrix and $R$ is the classical $R$-transform of the Hermitization
\begin{equation}\label{hermitianise}
\mathbb{B} =
\begin{pmatrix}
0 & \mathbf{B}\\
\mathbf{B}^\ast & 0
\end{pmatrix},
\end{equation}
equivalently, of the symmetrized singular-value distribution of $\mathbf{B}$. This leads to the \emph{single ring theorem}, relating the complex eigenvalue and singular value distributions—first obtained in the physics literature using diagrammatic methods~\cite{feinberg1997non}, later interpreted within free probability~\cite{haagerup2000brown}, and finally established rigorously for large bi-invariant random matrices~\cite{guionnet2011single}.

Beyond these special cases, there is no universally simple recipe for the $2\times2$ matrix-valued $\mathcal{R}$-transform of a given ensemble. Our contribution is to propose an alternative route to its computation based on spherical integrals. Let us briefly review their definition.

The Harish--Chandra--Itzykson--Zuber (HCIZ) integral is
\begin{equation}\label{eq:hciz}
\mathcal{I}_N(\mathbf{A},\mathbf{B};\theta)
\;=\;
\int_{\mathrm{U}(N)}
\exp\!\bigl(\theta\,\Tr(\mathbf{B}\,\mathbf{U}\,\mathbf{A}\,\mathbf{U}^{*})\bigr)\,\mathrm{d}\mathbf{U},
\end{equation}
where $\mathbf{A}$ and $\mathbf{B}$ are Hermitian matrices of size $N$, and $d\mathbf{U}$ is the Haar measure on the unitary group. 
Exact evaluations trace back to Harish--Chandra and Itzykson--Zuber~\cite{HarishChandra1957,ItzyksonZuber1980}; see also~\cite{ZinnJustinZuber2003}. We are interested here in the large-$N$ limit and in the spiked case. 
A particularly transparent result shows that if the spectral measure of $\mathbf{A}_N$
converges to a compactly supported law $\mu$, then for small~$\theta$
\begin{equation}\label{defH}
\frac{1}{N}\log 
\mathbb{E}_{\mathbf{U}}\!\Bigl[
\exp\!\bigl(\Tr(\mathbf{F}_N\,\mathbf{U}\,\mathbf{A}_N\,\mathbf{U}^{*})\bigr)
\Bigr]
\mathop{\longrightarrow}\limits_{N \to \infty}
\int_{0}^{\theta} R_{\mu}(t)\,\mathrm{d}t,
\end{equation}
with $\mathbf{F}_N=\mathrm{diag}(\theta,0,\dots,0)$, thereby identifying the
$R$-transform as the log-cumulant governing additive free convolution
\cite{MarinariParisiRitort1994,guionnet2005fourier}.
 If $\theta$ is sufficiently close to $0$, and if $\mathbf{A}$ is Hermitian and rotationally invariant\footnote{In the sense that $\mathbf{M}$ and $\mathbf{U}\mathbf{M}\mathbf{U}^*$ have the same distribution for every unitary matrix $\mathbf{U}$.}, the expectation over $\mathbf{U}$ may be replaced by an expectation over $\mathbf{A}$ alone; see~\cite{potters2020first} for an extended discussion. 
Consequently, the $R$-transform of a rotationally invariant matrix depends only on the law of a single diagonal entry of~$\mathbf{A}$ (see~\cite{guionnet2005fourier}).

In the non-Hermitian case, only the bi-invariant setting has been studied by Benaych-Georges~\cite{benaych2011rectangular}. He shows that
\begin{equation}
\frac{1}{2N}\log \E_{\vb{U}, \vb{V}}\!\Bigl[\exp\!\bigl(\Tr(F \vb{U}\vb{B}\vb{V})\bigr)\Bigr]
\underset{N\to +\infty}{\longrightarrow}
\int_{0}^{\theta} R(t)\,\mathrm{d}t
\end{equation}
with $F_N=\mathrm{diag}(\theta,0,\dots,0)$ (and, in fact, in a more general rectangular framework), where $R$ denotes the $R$-transform of the symmetrized law of the singular values of $\vb{B}$.

We would like to generalize these spherical integrals to the general case of large non-Hermitian matrices. Here is the new definition.

\begin{definition*}\label{def:H}
Let $\mathbf{M}$ be a deterministic or random matrix of size $N\times N$,
and let $\psi_1,\psi_2\in\mathbb C^N$. We define
$\mathcal H_{\mathbf M}^N$ by
\begin{equation}
\frac{1}{2N}
\log
\E_{\mathbf M,\mathbf U}\!\left[
\exp\!\Bigl(
2N\,\Re\langle
\psi_1,\mathbf U\mathbf M\mathbf U^*\psi_2
\rangle
\Bigr)
\right]
=
\mathcal H_{\mathbf M}^N\!\left(
\|\psi_1\|\|\psi_2\|,
\langle\psi_1,\psi_2\rangle,
\overline{\langle\psi_1,\psi_2\rangle}
\right).
\end{equation}\footnote{Here $\langle \cdot , \cdot \rangle$ denotes the standard inner product on $\C^N$,
$\langle x,y\rangle = \sum_{i=1}^N \overline{x_i}\,y_i$.}
where the expectation $\E_{\mathbf M,\mathbf U}$ is taken with respect to the randomness of $\mathbf M$ (when $\mathbf M$ is random) and to an independent Haar-distributed unitary random matrix $\mathbf U$.
We then set
\begin{equation}
\mathcal{H}_{\mathbf{M}} \;=\; \lim_{N\to\infty} \mathcal{H}_{\mathbf{M}}^N.
\end{equation}
Furthermore, we define the derivatives
\begin{equation}\label{R_transforms_def}
\mathcal{R}_1(\alpha,\beta, \overline{\beta}) \;:=\; \partial_{\alpha}\,\mathcal{H}(\alpha,\beta, \overline{\beta}),
\qquad
\mathcal{R}_2(\alpha,\beta, \overline{\beta}) \;:=\; 2\partial_{\beta}\,\mathcal{H}(\alpha,\beta, \overline{\beta})\footnote{%
The derivative $\partial_{\beta}$ denotes the Wirtinger derivative, defined as
$\displaystyle \partial_{\beta} = \tfrac{1}{2}\bigl(\partial_{\Re(\beta)} - i\,\partial_{\Im(\beta)}\bigr)$.
}
\end{equation}
We also denote by $\widetilde{\mathcal{R}}_1$ and $\widetilde{\mathcal{R}}_2$ the multivalued analytic functions 
that collect all possible determinations (branches) of $\mathcal{R}_1$ and $\mathcal{R}_2$, respectively. We say that the functions $\mathcal{R}_1$ and $\mathcal{R}_2$ defined by Eqs.~\eqref{R_transforms_def} are the principal branches of $\widetilde{\mathcal{R}}_1$ and $\widetilde{\mathcal{R}}_2$.
\end{definition*}

\begin{remarks}
    \item From this definition, in particular, if $\mathbf{A}$ and $\mathbf{B}$ are independent random (or deterministic) matrices and 
$\mathbf{U}$ is a Haar-distributed unitary random matrix, then
\begin{equation}
\mathcal{H}_{\mathbf{M}} \;=\; \mathcal{H}_{\mathbf{A}} + \mathcal{H}_{\mathbf{B}},
\qquad 
\text{where}\quad \mathbf{M} \;=\; \mathbf{A} + \mathbf{U}\mathbf{B}\mathbf{U}^*.
\end{equation}
Hence,
\begin{equation}
\mathcal{R}_{i,\mathbf{M}} \;=\; \mathcal{R}_{i,\mathbf{A}} + \mathcal{R}_{i,\mathbf{B}},
\qquad i\in\{1,2\},
\end{equation}
and, at the level of multivalued extensions,
\begin{equation}
\widetilde{\mathcal{R}}_{i,\mathbf{M}} \;=\; \widetilde{\mathcal{R}}_{i,\mathbf{A}} + \widetilde{\mathcal{R}}_{i,\mathbf{B}},
\qquad i\in\{1,2\}.
\end{equation}

\item In the case where $\vb{M}$ is Hermitian, one is led to consider the quantity
\begin{equation}
    \frac{1}{2N}\log 
    \mathbb{E}_{\mathbf{U}}\!\Bigl[
        \exp\!\bigl(\Tr(\mathbf{F}_N\,\mathbf{U}\,\mathbf{M}_N\,\mathbf{U}^{*})\bigr)
    \Bigr],
\end{equation}
where
\begin{equation}
    \vb{F}_N
    =
    \ket{\psi_1}\bra{\psi_2}
    +
    \ket{\psi_2}\bra{\psi_1}.
\end{equation}
We can then use the finite-rank HCIZ formula for $\vb{M}$ by introducing
the two eigenvalues $\lambda_{+}$ and $\lambda_{-}$ of $\vb{F}_N$.
Let
\begin{equation}
    \lambda_{\pm}
    =
    \Re(\beta)
    \pm
    \sqrt{\alpha^2-\Im(\beta)^2}.
\end{equation}
Then
\begin{equation}
    \mathcal{H}_{\vb{M}}(\alpha,\beta)
    =
    \frac{1}{2}\left(
        H_{\vb{M}}(\lambda_+)
        +
        H_{\vb{M}}(\lambda_-)
    \right),
\end{equation}
where $H_{\vb{M}}$ is the Hermitian transform defined in
Equation~\eqref{defH} by
\begin{equation}
    R_{\vb{M}}(x) = H'_{\vb{M}}(x).
\end{equation}

Thus, we obtain
\begin{equation}
    \mathcal{R}_{1,\vb{M}}(\alpha,\beta)
    =
    \frac{\alpha}{2\sqrt{\alpha^2-\Im(\beta)^2}}
    \left[
        R_{\vb{M}}(\lambda_+)
        -
        R_{\vb{M}}(\lambda_-)
    \right],
\end{equation}
and
\begin{equation}
    \mathcal{R}_{2,\vb{M}}(\alpha,\beta)
    =
    \frac{1}{2}\left(
        1
        +
        i\frac{\Im(\beta)}{\sqrt{\alpha^2-\Im(\beta)^2}}
    \right)
    R_{\vb{M}}(\lambda_+)
    +
    \frac{1}{2}\left(
        1
        -
        i\frac{\Im(\beta)}{\sqrt{\alpha^2-\Im(\beta)^2}}
    \right)
    R_{\vb{M}}(\lambda_-).
\end{equation}
\end{remarks}

We can now state the main result of this paper, obtained through the replica method. 
A brief overview of this method is given in Section~\ref{sec:replica} (Appendix), 
and the derivation of the result is discussed in Section~\ref{sec:derivation}.

\begin{result}\label{main_result}
Let $\vb{A}$ and $\vb{B}$ be two large independent random matrices, and let $\vb{U}$ be an independent Haar-distributed unitary random matrix. Then, for large $z\in\C$ or large $\omega\in \R$, one expects
\begin{equation}
    \E_{\vb{B}, \vb{U}}\!\left[\mathcal{G}_{\vb{A} + \vb{U}\vb{B}\vb{U}^*}(\omega, z)\right] 
    \;=\; \mathcal{G}_{\vb{A}}\!\left(\omega - \mathcal{R}_{1, \vb{B}}(\g_1(\omega, z),\g_2(\omega, z)),\, z - \mathcal{R}_{2, \vb{B}}(\g_1(\omega, z),\g_2(\omega, z))\right).
\end{equation}
This identity can be analytically continued to all $(\omega,z)$, although one must carefully choose the appropriate branches of $\widetilde{\mathcal{R}_1}$ and $\widetilde{\mathcal{R}_2}$.
\end{result}

\begin{remarks}
    \item By taking the limit $\omega \to 0$, one finds in particular that the $2\times2$ $\mathcal{R}$-transform defined in Eq.~\eqref{eq}
can be expressed in terms of $\mathcal{R}_1$ and $\mathcal{R}_2$ (and thus of the $\mathcal{H}$-transform) as
\begin{equation}
\mathcal{R}(\g_1,\g_2)
\;=\;
\begin{pmatrix}
    \mathcal{R}_1(\g_1,\g_2) & \mathcal{R}_2(\g_1,\g_2) \\
    \overline{\mathcal{R}_2(\g_1,\g_2)} & \mathcal{R}_1(\g_1,\g_2)
\end{pmatrix}.
\end{equation}
\end{remarks}

In the case $\vb{A}=0$, several interesting consequences follow. The result can become

\begin{result}\label{consequence}
    Let $\vb{B}$ be a large rotationally invariant random matrix.  Then, for large $\omega$ and $z\in\C$, one expects
\begin{equation}\label{main1}
    \g_{1, \vb{B}} = \frac{\omega - \mathcal{R}_{1, \vb{B}}(\g_1, \g_2)}{\left(\omega - \mathcal{R}_{1, \vb{B}}(\g_1, \g_2)\right)^2 - \left|z - \mathcal{R}_{2, \vb{B}}(\g_1, \g_2)\right|^2},
\end{equation}
\begin{equation}\label{main2}
    \overline{\g_{2, \vb{B}}} = \frac{\mathcal{R}_{2, \vb{B}}(\g_1, \g_2) - z}{\left(\omega - \mathcal{R}_{1, \vb{B}}(\g_1, \g_2)\right)^2 - \left|z - \mathcal{R}_{2, \vb{B}}(\g_1, \g_2)\right|^2}.
\end{equation}
This identity can be analytically continued to all $(\omega,z)$, although one must carefully choose the appropriate branches of $\mathcal{R}_1$ and $\mathcal{R}_2$.
\end{result}

\begin{remarks}
    \item If $\vb{M}$ is an $N\times N$ rotationally invariant random matrix, then
\begin{equation}\label{eq:joint_law}
\mathcal{H}_{\vb{M}}^N(\alpha,\beta)
= \frac{1}{2N}\log\E\!\Bigl[\exp\bigl(2N\bigl(\Re(\beta M_{11})
+\sqrt{\alpha^2-|\beta|^2}\,\Re M_{12}\bigr)\bigr)\Bigr].
\end{equation}
This shows something rather interesting: the Green functions of a rotationally invariant matrix $\vb{M}$ depend only on the joint law of one diagonal entry and one off-diagonal entry.
\item Interestingly, Eqs.~\eqref{main1} and~\eqref{main2} can be used to define possible expressions for 
$\mathcal{R}_1$ and $\mathcal{R}_2$ as
\begin{equation}\label{functionR1}
    \mathcal{R}_1(\g_1, \g_2)
    \;=\;
    \omega(\g_1, \g_2)
    \;-\;
    \frac{\g_1}{\g_1^2-|\g_2|^2},
\end{equation}
and
\begin{equation}\label{functionR2}
    \mathcal{R}_2(\g_1, \g_2)
    \;=\;
    z(\g_1, \g_2)
    \;-\;
    \frac{\overline{\g_2}}{|\g_2|^2 - \g_1^2},
\end{equation}
where $(\omega, z)(\g_1, \g_2)$ denote the functional inverses of $(\g_1, \g_2)$ with the domain restricted to the region where $\g_1^2\in \R$. Although these formulas offer an appealing parallel with the Hermitian case (see~\cite{potters2020first}),
they are practically unusable due to the difficulty of determining the inverse functions.

\end{remarks}

An interesting limit is obtained by setting $\omega = -i\epsilon$ with $\epsilon>0$ and $\epsilon \to 0$. In this regime, we introduce the function $\oo(z)$ such that the limit of $\g_1$ can be written as $i\,\oo(z)$. 
Moreover, $\g_2(z)$ converges to the Stieltjes transform of $\vb{B}$. The equations then take the form stated in the following result.

\begin{result}
    Let $\vb{B}$ be a large rotationally invariant random matrix.  Then, for $z$ inside the spedctrum, one expects

    \begin{equation}\label{main11}
    i\,\oo(z)
    \;=\;
    \frac{-\,\mathcal{R}_{1, \vb{B}}(i\,\oo(z), \g(z))}
    {\mathcal{R}_{1, \vb{B}}(i\,\oo(z), \g(z))^2
    - \left|z - \mathcal{R}_{2, \vb{B}}(i\,\oo(z), \g(z))\right|^2},
\end{equation}
\begin{equation}\label{main22}
    \overline{\g(z)}
    \;=\;
    \frac{\mathcal{R}_{2, \vb{B}}(i\,\oo(z), \g(z)) - z}
    {\mathcal{R}_{1, \vb{B}}(i\,\oo(z), \g(z))^2
    - \left|z - \mathcal{R}_{2, \vb{B}}(i\,\oo(z), \g(z))\right|^2}.
\end{equation}
\end{result}

\begin{remarks}
    \item Solving this system yields $\g(z)$, and therefore the spectral density $\rho(z)$ via Gauss’s law, 
$\partial_z \g = \pi \rho$. One can read in~\cite{janik1999correlations} that 
$\frac{\oo_{\vb{M}}(z)\oo_{\vb{M}^*}(\overline{z})}{\pi}$ is related to the so-called 
\emph{self-overlap} evaluated at $z$, introduced in~\cite{chalker1998eigenvector}.
This quantity measures the non-normality of eigenvectors and remains an active area of research. The functions $\mathcal{R}_1$ and $\mathcal{R}_2$ encode essential information not only about the distribution of complex eigenvalues and eigenvectors, but also about the singular values. Indeed, when $z = 0$, $\g_1$ reduces to the resolvent of $\vb{B}$, whose imaginary part yields the singular-value distribution. 
Thus, $\mathcal{R}_1$ and $\mathcal{R}_2$ together encode both singular-value and eigenvalue statistics. 
The relation between these two is generally nontrivial due to the coupled structure of the equations, 
except in the bi-invariant case where all expressions simplify, as mentioned in the introduction. All procedures described in this remark will be applied to the elliptic case 
in Appendix~\ref{sec:elliptic}.
\end{remarks}

One may also study equations~\eqref{main1} and~\eqref{main2} outside the spectrum, by taking \(z\) outside the support of the limiting spectral distribution and then letting \(\omega \to 0\). This is the content of the following result.

\begin{result}
Let $\vb{B}$ be a large rotationally invariant random matrix. For $z$ outside the spectral support of $\vb{B}$, define
\begin{equation}
    h_{\vb{B}}(z)
    =
    \tau\!\left(
    \left[
    (z\vb{1}-\vb{B})(z\vb{1}-\vb{B})^*
    \right]^{-1}
    \right).
\end{equation}

Then, for each connected component $\Omega$ of the complement of the spectral support of $\vb{B}$, there are two possible cases.

Either there exist branches $\mathcal{R}_{1,\vb{B}}$ and $\mathcal{R}_{2,\vb{B}}$ such that, for all $z\in\Omega$,
\begin{equation}\label{eq23}
    \partial_\alpha \mathcal{R}_{1,\vb{B}}(0,\g_{\vb{B}}(z))
    =
    \frac{1}{|\g_{\vb{B}}(z)|^2}
    -
    \frac{1}{h_{\vb{B}}(z)},
\end{equation}
and
\begin{equation}\label{eq24}
    \g_{\vb{B}}(z)
    =
    \frac{1}{z-\mathcal{R}_{2,\vb{B}}(0,\g_{\vb{B}}(z))}.
\end{equation}
On the unbounded connected component, and in particular when $|z|\to+\infty$, the branches $\mathcal{R}_{1,\vb{B}}$ and $\mathcal{R}_{2,\vb{B}}$ are precisely the principal branches.

Or else there exists a branch $\mathcal{R}_{1,\vb{B}}$ such that, for all $z\in\Omega$,
\begin{equation}
    \mathcal{R}_{1,\vb{B}}\!\left(\alpha,-(z-\tau(\vb{B}))\alpha^2\right)
    =
    -\frac{1}{\alpha}
    -
    \left(
    |z-\tau(\vb{B})|^2
    -
    \frac{1}{h_{\vb{B}}(z)}
    \right)\alpha
    +
    o(\alpha),
\end{equation}
and
\begin{equation}
    \g_{\vb{B}}(z)=0
\end{equation}
for all $z\in\Omega$. The second case can therefore occur only on a bounded simply connected component $\Omega$.
\end{result}

\begin{remarks}
    \item On each domain $\Omega$, the relevant branches of
    $g\mapsto \partial_\alpha \mathcal{R}_{1,\vb{B}}(0,g)$ and
    $g\mapsto \mathcal{R}_{2,\vb{B}}(0,g)$ are obtained by analytic continuation. Thus, knowing one branch on a given domain $\Omega$ may be enough to recover the other possible branches on the same domain. However, knowing the transforms on one component gives no information about the transforms on the other components.

    \item For the principal branches $\mathcal{R}_{1,\vb{B}}$ and $\mathcal{R}_{2,\vb{B}}$, an asymptotic analysis of Eqs.~\eqref{eq23} and~\eqref{eq24} as $|z|\to+\infty$ gives
\begin{equation}
    \mathcal{R}_{2,\vb{B}}(0,g)
    =
    \sum_{k=1}^{+\infty}
    \kappa_k(\vb{B})g^{k-1},
\end{equation}
where the $\kappa_k(\vb{B})$ are the usual free cumulants, as defined for instance in~\cite{potters2020first}. These cumulants are additive under free convolution. In particular,
\begin{equation}
    \kappa_1(\vb{B})=\tau(\vb{B}),
    \qquad
    \kappa_2(\vb{B})=\tau(\vb{B}^2)-\tau(\vb{B})^2.
\end{equation}

Similarly, the expansion of $\partial_\alpha\mathcal{R}_{1,\vb{B}}(0,g)$ defines another family of additive cumulants:
\begin{equation}
    \partial_\alpha\mathcal{R}_{1,\vb{B}}(0,g)
    =
    \sum_{k=2}^{+\infty}
    \eta_k(\vb{B})g^{k-2}.
\end{equation}
The first one is
\begin{equation}
    \eta_2(\vb{B})
    =
    \tau(\vb{B}\vb{B}^*)-|\tau(\vb{B})|^2.
\end{equation}
    \item The last sentence of the result follows from a simple observation. Consider a domain $\Omega$ on which $\g_{\vb{B}}(z)=0$. Then no loop in $\Omega$ can enclose any charge, by Gauss's theorem. It follows that $\Omega$ cannot contain a hole. Moreover, $\Omega$ is necessarily bounded, since
    \begin{equation}
        \g_{\vb{B}}(z)
        \underset{|z|\to+\infty}{\sim}
        \frac{1}{z}.
    \end{equation}
\end{remarks}
To summarize what this approach could bring—and what will be the subject of future work—note that the two scalar functions $\mathcal{R}_1(\alpha,\beta)$ and $\mathcal{R}_2(\alpha,\beta)$ both arise from a single function $\mathcal{H}$. We hope this will provide a new class of tractable examples. The emergence of spherical integrals was previously known only in the rectangular bi-invariant case, due to Benaych-Georges \cite{benaych2011rectangular}. The result is based on the powerful—but non-rigorous—replica method, which has been extremely successful in various contexts (e.g., random matrix theory and disordered systems; see \cite{mezard1987spin}). In recent works (e.g.\ \cite{belinschi2015operator,kargin2015subordination,bercovici2022brown,belinschi2018eigenvalues}), subordination is often expressed via analytic maps $\omega_1,\omega_2$ defined by
\begin{equation}
  \omega_1(z) \;=\; \mathcal{R}_1\bigl(i\,\oo(z),\,\g(z)\bigr),
  \qquad
  \omega_2(z) \;=\; \mathcal{R}_2\bigl(i\,\oo(z),\,\g(z)\bigr).
\end{equation}
Rather than working with the analytic functions $\omega_1$ and $\omega_2$, we prefer to use $\mathcal{R}_1$ and $\mathcal{R}_2$ directly. Writing $\omega_1$ and $\omega_2$ in this way, however, forces one to regard $\mathcal{R}_1$ and $\mathcal{R}_2$ as multivalued functions. The result may therefore offer a promising framework for studying the complex spectrum of deformed matrices, leading to general results that extend what is known about the Brown measure of deformations of the Ginibre matrix, the elliptic matrix, the Hermitian matrix, or the bi-invariant matrix \cite{haagerup2000brown,bercovici2022brown,HallHo2022ImagSemicirc,Ho2022SelfAdjointPlusElliptic,HoYinZhong2025OutliersFullRankSingleRing,HoZhong2025DeformedSingleRing,Zhong2021EllipticPlusFree}. These are precisely the cases in which $\mathcal{R}_1$ and $\mathcal{R}_2$ are under control. The general case will require treating these functions as multivalued, which is likely to lead to a difficult theory. Likewise, one can hope to study the singular values of $\vb{A}+\vb{B}$. The rectangular case for bi-invariant matrices follows from the work of Florent Benaych-Georges \cite{benaych2009rectangular}. Here, we have a framework that can handle the case where $\vb{B}$ is merely rotationally invariant.

\paragraph*{Acknowledgements.} 

We are grateful to Florent Benaych-Georges, Jean-Philippe Bouchaud, Roland Speicher and Zdzisław Burda for their valuable insights. This research was conducted within the Econophysics \& Complex Systems Research Chair, under the aegis of the Fondation du Risque, the Fondation de l’Ecole polytechnique, the Ecole polytechnique, and Capital Fund Management.

\bibliographystyle{unsrt}

\bibliography{References.bib}

\newpage
\appendix

\section{APPENDICES}
\subsection{Reminder on the replica method for random matrices}
\label{sec:replica}
Let $\vb{C}$ be an $N\times N$ Hermitian matrix with strictly positive real eigenvalues. The standard complex Gaussian identities yield
\begin{equation}
    \frac{1}{\pi^N}
    \int_{\C^N}
    \phi_a\,\overline{\phi_b}\,
    e^{-\langle \phi,\vb{C}\phi\rangle}\,
    \mathrm{d}\phi
    \;=\;
    \frac{[\vb{C}^{-1}]_{ab}}{\det(\vb{C})},
    \qquad 1\leq a,b\leq N,
\end{equation}
where $d\phi$ denotes the Lebesgue measure on $\C^N$.
As a consequence,
\begin{equation}
    [\vb{C}^{-1}]_{ab}
    \;=\;
    \frac{\displaystyle
    \int_{\C^N}
    \phi_a\,\overline{\phi_b}\,
    e^{-\langle \phi,\vb{C}\phi\rangle}\,
    d\phi}
    {\displaystyle
    \int_{\C^N}
    e^{-\langle \phi,\vb{C}\phi\rangle}\,
    \mathrm{d}\phi}.
\end{equation}

To introduce replicas, we multiply both the numerator and the denominator by
\(
\bigl(\pi^{-N}\!\int_{\C^N} e^{-\langle \phi,\vb{C}\phi\rangle}\,\mathrm{d}\phi\bigr)^{n-1},
\)
and introduce $n$ independent copies
$\phi^{(1)},\dots,\phi^{(n)}\in\C^N$.
This gives
\begin{equation}
    [\vb{C}^{-1}]_{ab}
    \;=\;
    \det(\vb{C})^{\,n}
    \int_{(\C^N)^n}
    [\phi^{(1)}]_a\,\overline{[\phi^{(1)}]_b}\,
    \exp\!\Bigl(
        -\sum_{k=1}^n
        \langle \phi^{(k)},\vb{C}\phi^{(k)}\rangle
    \Bigr)\,
    \frac{\mathrm{d}\phi^{(1)}\cdots \mathrm{d}\phi^{(n)}}{\pi^{Nn}}.
\end{equation}
This identity holds for all integers $n\ge 1$.
The replica method then consists in formally taking the limit $n\to 0$,
using $\det(\vb{C})^{\,n}\to 1$, which leads to
\begin{equation}\label{eq:replica}
    [\vb{C}^{-1}]_{ab}
    \;=\;
    \lim_{n\to 0}
    \int_{(\C^N)^n}
    [\phi^{(1)}]_a\,\overline{[\phi^{(1)}]_b}\,
    \exp\!\Bigl(
        -\sum_{k=1}^n
        \langle \phi^{(k)},\vb{C}\phi^{(k)}\rangle
    \Bigr)\,
    \frac{\mathrm{d}\phi^{(1)}\cdots \mathrm{d}\phi^{(n)}}{\pi^{Nn}}.
\end{equation}

Formula~\eqref{eq:replica} provides a convenient representation for resolvent entries in Hermitian settings.
It is particularly well suited for saddle-point analysis and behaves naturally under sums of independent matrices.

We now consider a $2N\times 2N$ Hermitian matrix $\vb{C}$ with block structure
\begin{equation}
    \vb{C} \;=\;
    \begin{pmatrix}
        \vb{C}_{11} & \vb{C}_{12} \\
        \vb{C}_{12}^* & \vb{C}_{22}
    \end{pmatrix},
    \qquad
    \vb{C}_{11},\vb{C}_{22}, \vb{C}_{12}\in\C^{N\times N}.
\end{equation}
Writing the inverse in block form,
\begin{equation}
    \vb{C}^{-1} \;=\;
    \begin{pmatrix}
        \vb{G}_{11} & \vb{G}_{12} \\
        \vb{G}_{21}^* & \vb{G}_{22}
    \end{pmatrix},
\end{equation}
the replica representation yields, for $i,j\in\{1,2\}$,
\begin{multline}\label{eq:block-replica}
    \vb{G}_{ij}
    \;=\;
    \lim_{n\to 0}
    \int_{\phi_1^{(k)},\,\phi_2^{(k)}\in\C^N}
    \phi_i^{(1)}\,\left(\phi_j^{(1)}\right)^*
    \\
    \times
    \exp\!\Biggl(
        -\sum_{k=1}^n\Bigl[
            \langle \phi_1^{(k)},\vb{C}_{11}\phi_1^{(k)}\rangle
            + \langle \phi_2^{(k)},\vb{C}_{22}\phi_2^{(k)}\rangle
            + 2\,\Re\!\bigl\langle \phi_1^{(k)},\vb{C}_{12}\phi_2^{(k)}\bigr\rangle
        \Bigr]
    \Biggr)
    \,
    \frac{\mathrm{d}\phi_1^{(1)}\!\cdots \mathrm{d}\phi_1^{(n)}\,
          \mathrm{d}\phi_2^{(1)}\!\cdots \mathrm{d}\phi_2^{(n)}}{\pi^{2Nn}}.
\end{multline}
We now specialize to the matrix
\begin{equation}
    \vb{C} =
    \begin{pmatrix}
        \omega\,\vb{I}_N & z\,\vb{I}_N - \vb{M} \\
        (z\,\vb{I}_N - \vb{M})^* & \omega\,\vb{I}_N
    \end{pmatrix},
\end{equation}
where $z\in\C$, $\vb{M}$ is a non-Hermitian $N\times N$ matrix, and $\omega>0$. We assume that either $\omega$ or $|z|$ is large, so that $\vb{C}$ is a Hermitian matrix with strictly positive eigenvalues. In this case, the representation~\eqref{eq:block-replica} becomes
\begin{multline}\label{bigformula}
    \vb{G}_{ij}(\omega,z)
    \;=\;
    \lim_{n\to 0}
    \int_{\phi_1^{(k)},\,\phi_2^{(k)}\in\C^N}
    \phi_i^{(1)}\,\left(\phi_j^{(1)}\right)^*
    \\
    \times
    \exp\!\Biggl(
        -\sum_{k=1}^n\Bigl[
            \omega \|\phi_1^{(k)}\|^2
            + \omega \|\phi_2^{(k)}\|^2
            + 2\,\Re\!\bigl\langle \phi_1^{(k)},(z-\vb{M})\phi_2^{(k)}\bigr\rangle
        \Bigr]
    \Biggr)
    \,
    \frac{\mathrm{d}\phi_1^{(1)}\!\cdots \mathrm{d}\phi_1^{(n)}\,
          \mathrm{d}\phi_2^{(1)}\!\cdots \mathrm{d}\phi_2^{(n)}}{\pi^{2Nn}}.
\end{multline}

\subsection{Derivation of the result using the replica trick}\label{sec:derivation}

Our approach is very similar to the one used in the Hermitian case \cite{bun2016rotational}. 
To simplify notation, we work with a single replica $n=1$ and absorb the overall exponential factor at the end.

We apply formula~\eqref{bigformula} in the case $\vb{M} = \vb{A} + \vb{U}\vb{B}\vb{U}^*$:
\begin{multline}
    [\vb{G}_{\vb{M}}(\omega, z)]_{i j} 
    \;\propto\;
    \int_{\phi_1,\,\phi_2 \in \C^N} 
        \phi_i \,\phi_j^* \,
        \exp\!\left\{-\omega \|\phi_1\|^2 \;-\; \omega \|\phi_2\|^2 \;-\; 2\,\Re\!\bigl(z\,\langle \phi_1, \phi_2\rangle\bigr)\right\} \\
    \times 
        \exp\!\left\{2\,\Re\!\bigl(\langle \phi_1, \vb{A}\,\phi_2\rangle\bigr)\right\}
        \exp\!\left\{2\,\Re\!\bigl(\langle \phi_1, \vb{U}\vb{B}\vb{U}^*\,\phi_2\rangle\bigr)\right\}
    \,\mathrm{d}\phi_1 \,\mathrm{d}\phi_2 .
\end{multline}
Now take expectation with respect to $\vb{U}$ using the definition of $\mathcal{H}_{\vb{B}}$ in Definition~\ref{def:H}:
\begin{multline}
    \E_{\vb{B}, \vb{U}}\!\bigl([\vb{G}_{\vb{M}}(\omega, z)]_{i j}\bigr) 
    \;\propto\;
    \int_{\phi_1,\,\phi_2 \in \C^N} 
        \phi_i \,\phi_j^* \,
        \exp\!\left\{-\omega \|\phi_1\|^2 \;-\; \omega \|\phi_2\|^2 \;-\; 2\,\Re\!\bigl(z\,\langle \phi_1, \phi_2 \rangle\bigr)\right\} \\
    \times 
        \exp\!\left\{2\,\Re\!\bigl(\langle \phi_1, \vb{A}\,\phi_2\rangle\bigr)\right\}
        \exp\!\left\{2N \,\mathcal{H}_{\vb{B}}\!\Bigl(
            \tfrac{\|\phi_1\|}{\sqrt{N}}\;\tfrac{\|\phi_2\|}{\sqrt{N}},\;
            \tfrac{\langle \phi_1, \phi_2\rangle}{N}
        \Bigr)\right\}
    \,\mathrm{d}\phi_1 \,\mathrm{d}\phi_2 .
\end{multline}
Introduce the Lagrange multipliers:
\begin{equation}
    C \;=\; \frac{\|\phi_1\|^2}{N}, 
    \qquad 
    D \;=\; \frac{\|\phi_2\|^2}{N}, 
    \qquad 
    E \;=\; \frac{\langle \phi_1, \phi_2\rangle}{N},
\end{equation}
and recall the representation of the Dirac delta as an integral over the imaginary axis:
\begin{equation}
    \delta(x)
    \;=\;
    \int_{-i\infty}^{i\infty} \frac{e^{-x y}}{2\pi i}\,dy.
\end{equation}
Then
\begin{multline}
    \E_{\vb{B}, \vb{U}}\!\bigl([\vb{G}_{\vb{M}}(\omega, z)]_{i j}\bigr) 
    \;\propto\;
    \int 
    \int_{\phi_1,\,\phi_2 \in \C^N} 
        \phi_i \,\phi_j^* \,
        \exp\!\left\{-\omega \|\phi_1\|^2 \;-\; \omega \|\phi_2\|^2 \;-\; 2\,\Re\!\bigl(z\,\langle \phi_1, \phi_2\rangle\bigr)\right\} \\
    \times 
        \exp\!\left\{2N \,\mathcal{H}_{\vb{B}}(\sqrt{C D},\,E)\right\} \\
    \times 
        \exp\!\left\{
            c\Bigl(\tfrac{\|\phi_1\|^2}{N} - C\Bigr)
            \;+\;
            d\Bigl(\tfrac{\|\phi_2\|^2}{N} - D\Bigr)
            \;+\;
            \Re\!\Bigl[\overline{e}\,\bigl(\tfrac{\langle \phi_1, \phi_2\rangle}{N} - E\bigr)\Bigr]
        \right\} \\
    \mathrm{d}\phi_1 \,\mathrm{d}\phi_2 \,
    \mathrm{d}c \,\mathrm{d}d \,\mathrm{d}\Re(e) \,\mathrm{d}\Im(e)\,\mathrm{d}C \,\mathrm{d}D \,\mathrm{d}\Re(E) \,\mathrm{d}\Im(E).
\end{multline}

Make the change of variables $c \mapsto cN$, $d \mapsto dN$, $e \mapsto 2Ne$:
\begin{multline}\label{ap}
    \E\!\bigl([\vb{G}_{\vb{M}}(\omega, z)]_{i j}\bigr) 
    \;\propto\;
    \int 
    \int_{\phi_1,\,\phi_2 \in \C^N} 
        \phi_i \,\phi_j^* \,
        \exp\!\left\{-(\omega - c)\|\phi_1\|^2 \;-\; (\omega - d)\|\phi_2\|^2 \;-\; 2\,\Re\!\bigl((z - \overline{e})\,\langle \phi_1, \phi_2\rangle\bigr)\right\} \\
    \times 
        \exp\!\left\{2N \,\mathcal{H}_{\vb{B}}(\sqrt{C D},\,E)\right\}
        \exp\!\left\{-N\,c\,C \;-\; N\,d\,D \;-\; 2N\,\Re\!\bigl[e\,E\bigr]\right\}
    \\\,\mathrm{d}\phi_1 \,\mathrm{d}\phi_2 \,
    \mathrm{d}c \,\mathrm{d}d \,\mathrm{d}\Re(e) \,\mathrm{d}\Im(e)\,\mathrm{d}C \,\mathrm{d}D \,\mathrm{d}\Re(E) \,\mathrm{d}\Im(E).
\end{multline}
We now perform the Gaussian integrals over $\phi_1$ and $\phi_2$. 
Equation~\eqref{ap} then yields
\begin{equation}\label{ap2}
    \E_{\vb{B}, \vb{U}}\!\bigl([\vb{G}_{\vb{M}}(\omega, z)]_{i j}\bigr)
    \;\approx\footnote{The symbol $\approx$ is inherently imprecise due to the use of the replica method. The result we aim to explain assumes that $\approx$ preserves the block operator $\mathcal{G}$ defined in Eq.~\eqref{defG}.}\; \bigl[\vb{G}_{\vb{A}}(\omega - c,\, z - \overline{e})\bigr]_{i j},
\end{equation}
where $c, d, e, C, D, E$ are chosen to optimize
\begin{equation}\label{eqaderivee}
    2\,\mathcal{H}(\sqrt{C D},\,E)
    \;-\;
    c\,C
    \;-\;
    d\,D
    \;-\;
    \overline{e}\,E
    \;-\;
    e\,\overline{E}
    \;+\;
    \log\!\bigl(\det \vb{A}\bigr).
\end{equation}
Taking derivatives of \eqref{eqaderivee} with respect to $C,\,D,\,E$ gives
\begin{equation}\label{system}
    \left\{
    \begin{aligned}
        \sqrt{\tfrac{D}{C}}\;\partial_{\alpha}\,\mathcal{H}_{\vb{B}}(\sqrt{C D},\,E) &= c, \\
        \sqrt{\tfrac{C}{D}}\;\partial_{\alpha}\,\mathcal{H}_{\vb{B}}(\sqrt{C D},\,E) &= d, \\
        2\,\partial_{\beta}\,\mathcal{H}_{\vb{B}}(\sqrt{C D},\,E) &= \overline{e}.
    \end{aligned}
    \right.
\end{equation}
Using the identity
\begin{equation}
    \partial_t \log\!\det\!\bigl(\vb{A}(t)\bigr)
    \;=\;
    \Tr\!\bigl(\vb{A}(t)^{-1}\,\vb{A}'(t)\bigr),
\end{equation}
the derivatives of \eqref{eqaderivee} with respect to $c,\,d,\,\overline{e}$ yield
\begin{equation}
    \left\{
    \begin{aligned}
        C = D &= \g_{1, \vb{M}}, \\
        E &= \g_{2, \vb{M}} ,
    \end{aligned}
    \right..
\end{equation}
Finally, system~\eqref{system} becomes
\begin{equation}
    \left\{
    \begin{aligned}
        c = d &= \partial_{\alpha}\,\mathcal{H}_{\vb{B}}\!\bigl(\g_{1, \vb{M}},\,\g_{2, \vb{M}}\bigr), \\
        \overline{e} &= 2\,\partial_{\beta}\,\mathcal{H}_{\vb{B}}\!\bigl(\g_{1, \vb{M}},\,\g_{2, \vb{M}}\bigr).
    \end{aligned}
    \right.
\end{equation}
Substituting these values of $c, d, e$ back yields the approximation
\begin{equation}\label{approx}
    \vb{G}_{\vb{M}}(\omega, z)
    \;\approx\;
    \vb{G}_{\vb{A}}\!\bigl(
        \omega \;-\; \mathcal{R}_{1, \vb{B}}(\g_1, \g_2),\;
        z \;-\; \mathcal{R}_{2, \vb{B}}(\g_1, \g_2)
    \bigr),
\end{equation}where $\mathcal{R}_1$ and $\mathcal{R}_{2}$ are the transforms defined in Eq.~\ref{R_transforms_def}.

\subsection{Example: the complex elliptic Ginibre ensemble}\label{sec:elliptic}

We now study a complete example within our framework: the complex elliptic Ginibre ensemble, denoted eGinUE. All results presented here are well known~\cite{sommers1988spectrum,girko1986elliptic,fyodorov1997almost,janik1999correlations},  
except for the statement concerning singular values, which, to the best of our knowledge, has not been explicitly noted before.  

These are $N \times N$ matrices with i.i.d.\ complex Gaussian entries of mean zero and correlations
\begin{equation}
\E|X_{ii}|^2] = \E[|X_{ij}|^2] = \frac{1}{N},
\qquad 
\E[X_{ij} \overline{X_{ji}}] = \tau,
\end{equation}
for all $1 \le i \ne j \le N$, with $\tau \in [-1, 1]$. The joint probability density function (JPDF) with respect to the Lebesgue measure  
$\prod_{i,j=1}^N dX_{ij}\,d\overline{X_{ij}}$ is
\begin{equation}\label{pdf_elliptic}
    \frac{1}{\pi^{N^2} (1-\tau^2)^{N^2/2}} 
    \exp\!\left[-\frac{1}{1 - \tau^2}\,
    \Tr(XX^* - \tau\,\Re X^2)\right] dX.
\end{equation}
This is a rotationally invariant ensemble. Using the JPDF~\eqref{pdf_elliptic} and Gaussian integration, one obtains that the $\mathcal{H}$-transform of this ensemble is
\begin{equation}
    \mathcal{H}(\alpha, \beta, \overline{\beta})
    \;=\;
    \frac{\alpha^2}{2}
    + \frac{\tau}{4}\bigl(\beta^2 + \overline{\beta}^2\bigr),
\end{equation}
and, using the definitions in Eq.~\eqref{R_transforms_def},
\begin{equation}
    \mathcal{R}_1(\alpha, \beta) = \alpha,
    \qquad
    \mathcal{R}_2(\alpha, \beta) = \tau\,\beta,
\end{equation}
which can also be derived diagrammatically~\cite{BurdaJanikNowakMultiplication}. Substituting into the general relations given by the result~\ref{consequence}, we obtain
\begin{equation}\label{main_equations_elliptic}
    \g_{1}
    \;=\;
    \frac{\omega - \g_1}
    {(\omega - \g_1)^2 - |z - \tau\,\g_2|^2},
\end{equation}
\begin{equation}
    \overline{\g_{2}}
    \;=\;
    \frac{\tau\,\g_2 - z}
    {(\omega - \g_1)^2 - |z - \tau\,\g_2|^2},
\end{equation}
for all $z$ and $\omega$, since $\mathcal{R}_1$ and $\mathcal{R}_2$ admit no other analytic continuations.

\paragraph{Eigenvalue distribution.}
Setting $\omega = 0$, it is straightforward to check that the Stieltjes transform is given by 
\begin{equation}\label{eqginside}
    \g(z)
    \;=\;
    \frac{\overline{z} - \tau\,z}{1 - \tau^2},
    \qquad z \in \Omega,
\end{equation}
and
\begin{equation}\label{eqgoutside}
    \g(z)
    \;=\;
    \frac{z - \sqrt{z^2 - 4\tau}}{2\tau},
    \qquad z \notin \Omega,
\end{equation}
where $\Omega$ is the filled ellipse
\begin{equation}
    \Omega
    \;=\;
    \left\{
    z \in \C :
    \frac{(\Re z)^2}{(1+\tau)^2}
    + \frac{(\Im z)^2}{(1 - \tau)^2}
    \le 1
    \right\}.
\end{equation}
Moreover,
\begin{equation}\label{eqoverlapelliptic}
    \oo(z)^2
    \;=\;
    \frac{1}{\pi}
    \left(
    1 - \frac{(\Re z)^2}{(1+\tau)^2}
      - \frac{(\Im z)^2}{(1 - \tau)^2}
    \right),
    \qquad z \in \Omega,
\end{equation}
and $\oo(z)^2 = 0$ otherwise.  
Using Gauss’s law, we then recover the limiting spectral density
\begin{equation}
    \rho(z)
    \;=\;
    \frac{\partial_{\overline{z}} \g(z)}{\pi}
    \;=\;
    \frac{1}{\pi(1 - \tau^2)}\,\mathbf{1}_{z \in \Omega}.
\end{equation}where $\mathbf{1}_{z \in \Omega}$ denotes the indicator function of the domain $\Omega$.

\paragraph{Singular-value distribution.}
Returning to Eqs.~\eqref{main_equations_elliptic} and setting $z = 0$,
the second relation gives $\g_2 = 0$, and the first yields
\begin{equation}
    \g_1 = \frac{1}{\omega - \g_1}.
\end{equation}
Here, $\g_1$ is the Stieltjes transform of the Hermitized matrix of $\vb{B}$, defined in Eq.~\eqref{hermitianise}. We see that it does not depend on $\tau$.  
For $\tau = 0$, one recovers the result of Marčenko and Pastur~\cite{marvcenko1967distribution},
corresponding to the quarter-circle law on $[0,2]$. Thus, the singular-value distribution of the elliptic Ginibre ensemble 
is independent of the parameter $\tau$ and follows the same quarter-circle law. This simple example clearly illustrates the profound difference between eigenvalue and singular-value statistics.

\end{document}